\documentclass[12pt]{iopart}
\usepackage{amsfonts}
\usepackage{mathrsfs}
\usepackage{graphicx}
\usepackage{wrapfig,floatrow}
\expandafter\let\csname equation*\endcsname\relax
\expandafter\let\csname endequation*\endcsname\relax
\usepackage{amsmath}
\usepackage{bbm}

\begin{document}

\title{Hidden Momentum and Black Hole Kicks}

\author{Samuel E. Gralla and Frank Herrmann}
\address{Department of Physics \\ University of Maryland \\
College Park, MD 20742-411 }

\begin{abstract}
A stationary magnetic dipole immersed in an electric field carries ``hidden'' mechanical momentum.  However, the fate of this momentum if the fields are turned off is unclear.  We consider a charge-and-dipole hidden momentum configuration, and turn off the fields by collapsing a null shell onto the system, forming a black hole.  In numerical calculations we find that the black hole receives a kick corresponding to $0.1\%$ of the initial stored momentum.  When extrapolated to apply to purely gravitational phenomena, this efficiency suggests a role for the hidden momentum kick mechanism in generating the binary black hole ``superkicks'' observed in numerical simulations of Einstein's equation.

\end{abstract}

\maketitle

\section{Background and Motivation}
First described nearly fifty years ago \cite{shockley-james}, the phenomenon of hidden mechanical momentum continues to generate interest \cite{hnizdo,babson-reynolds-bjorkquist-griffiths,gralla-harte-wald,costa-herdeiro-natario-zilhao,mcdonald,griffiths-hnizdo}. The basic idea is simple.  In flat spacetime, any time-independent, conserved stress-energy $T_{\mu \nu}$ must have zero total momentum $P^i = \int T^{i0} d^3 x$ (when this integral is defined).  For matter coupled to electromagnetic fields, we then have for stationary configurations that the total mechanical plus electromagnetic field momentum is zero,
\begin{equation}\label{ptot}
\vec{P}^M + \vec{P}^{EM} = 0.
\end{equation}
(Here the total stress energy $T_{\mu \nu}=T^M_{\mu \nu} + T^{EM}_{\mu \nu}$ is given by a matter component $T^M_{\mu \nu}$ and the field component $T^{EM}_{\mu \nu}$, and the momenta $P^M$ and $P^{EM}$ are defined in terms of their respective stress tensors.) However, using stationary sources it is easy to create stationary field configurations whose field momentum is non-zero.  Thus, the sources contain non-zero momentum---there is mechanical momentum in bodies at rest!  This is the hidden momentum.

The original and simplest example is a magnetic dipole immersed in an electric field.  Consider a point dipole of strength $\vec{\mu}$ located at a position $\vec{x}_{\textrm{dip}}$ in a stationary electric field $\vec{E}=-\vec{\nabla} \Phi$ assumed to vanish at infinity.  Then the field momentum is
\begin{equation}\label{pem}
\vec{P}^{EM} = \frac{1}{4\pi} \int \vec{E} \times \vec{B} = - \frac{1}{4\pi} \int \vec{\nabla} \Phi \times \vec{B} = \frac{1}{4\pi} \int \Phi \vec{\nabla} \times \vec{B} = \int \Phi \vec{J} = - \vec{\mu} \times \vec{E}|_{\vec{x}_{\textrm{dip}}}.
\end{equation}
The first equality is the definition of $\vec{P}^{EM}$, the second uses $\vec{E}=-\vec{\nabla} \Phi$, the third integrates by parts, the fourth uses the Maxwell equation $\vec{\nabla} \times \vec{B} = 4 \pi \vec{J}$,\footnote{In this paper we used Gaussian geometrized units, i.e., Gaussian units where additionally Newton's constant and the speed of light are set to unity.} and the fifth uses the point-dipole current $\vec{J}=-\vec{\mu} \times \vec{\nabla} \delta(\vec{x}-\vec{x}_{\textrm{dip}})$.  By equation \eqref{ptot}, the hidden mechanical momentum is then
\begin{equation}\label{pmech}
\vec{P}^M = \vec{\mu} \times \vec{E}|_{\vec{x}_{\textrm{dip}}}.
\end{equation}
Equation \eqref{pmech} gives the mechanical momentum present in a stationary dipole immersed in an electromagnetic field.  How is this momentum stored?  For a simple model of a current loop dipole, the electric field modifies the flow of current in such a way that the (relativistic) particle momenta of the charge carriers sums precisely to the simple formula above \cite{babson-reynolds-bjorkquist-griffiths,griffiths}.

The above discussion applies to stationary situations only, and it is natural to ask the fate of the hidden momentum if the fields are turned off or the configuration is otherwise destroyed by a time-dependent process.  Is some (or all) of the hidden mechanical momentum converted into ordinary momentum, resulting in a kick (balanced by emitted radiation)?  Or does the system just relax to the trivial configuration containing a body at rest?  The process is difficult to model, since one must include the agent turning off the fields as part of the calculation, and the result will in general depend on the details of the turning off process.  To our knowledge, no such calculation has yet been carried out.  A recent study of hidden momentum \cite{babson-reynolds-bjorkquist-griffiths} concluded that while some impulse may be delivered, ``there is no obvious reason why this impulse should equal the momentum originally stored in the fields---all we can say in general is that the total momentum afterward, like the total momentum before, is zero.''

In this paper we will determine the fraction of momentum that is converted into impulse in one particular scenario---collapse to a black hole.  Specifically, we will consider a point electric charge placed near a point magnetic dipole, both surrounded by a shell of matter assumed large and light enough that its gravity does not (initially) distort the electromagnetic field configuration.  Thus, initially, we have a perfectly ordinary hidden momentum configuration.  However, at a time $t_0$ the shell begins to collapse at the speed of light, beginning the process of black hole formation.  By the ``no hair theorem'' the final black hole can only support a simple monopolar electric field, so the final configuration has no hidden momentum and no field momentum.  By monitoring the momentum radiated  (relative to the rest frame of the black hole) we determine that a kick corresponding to $0.1\%$ of the initial stored momentum is delivered to the black hole.  To perform these calculations we adapt a model due to \cite{delacruz-chase-israel}, which treats the electromagnetic fields as small perturbations to the collapse spacetime and exploits causality to simplify the calculations.  Although the ``form a black hole'' method of turning off the fields may seem unduly complicated (or completely out of the blue), this type of trick makes it likely to be one of the \textit{simplest} situations to model.

A completely separate motivation for studying black hole formation in the presence of hidden momentum comes from numerical simulations of black hole mergers.  A surprising discovery \cite{campanelli-lousto-zlochower-merritt,campanelli-lousto-zlochower-merritt-prl,gonzalez-sperhake-burgmann-hannam-husa} was that for certain configurations of black hole binaries containing spinning black holes, there is a bobbing up and down of the entire orbital plane that correlates to a large kick velocity obtained at merger.  In previous work \cite{gralla-harte-wald}, this behavior was explained using the concept of hidden momentum.  By revealing the bobbing to be associated with a kinematical effect of spin that is present even for systems for which there can be no kick (an example is spinning balls connected by a string), the authors of \cite{gralla-harte-wald} concluded that the kick should not be viewed as an inertial continuation of the bobbing.  But what, then, causes the kick?  It was noted that, at least in an electromagnetic analog of orbiting charged magnetic dipoles, there is an orbitally modulated field momentum present outside the bodies, along with corresponding hidden momentum within the bodies.  The authors suggested that a rapid merger event allows the release of some of this field momentum, leading to a kick that will depend sinusoidally on the phase at merger, as observed in numerical simulations.  While this picture accounts qualitatively for the observed behavior, there was no quantitative estimate of the percentage of the field momentum that a merger could release.  Furthermore, there was no means of distinguishing the proposal from a competing scenario---also qualitatively successful---in which the kick is caused by the bobbing motion's modulation of the dominant orbital emission \cite{pretorius}.  The calculations of this paper provide an estimate of the efficiency of the momentum release mechanism in a context where the competing mechanism is absent, thereby helping to distinguish the two scenarios.  The implications for our understanding of black hole merger kicks are discussed in section \ref{sec:discussion}.

\section{Model and Results}\label{sec:model}

\begin{wrapfigure}{r}{0.4\textwidth}
  \vspace{-10pt} \hspace{-110pt}
  \ffigbox[290pt]
  {
      \caption{The collapse spacetime.}
  }
  {
    \includegraphics[width=.95\textwidth]{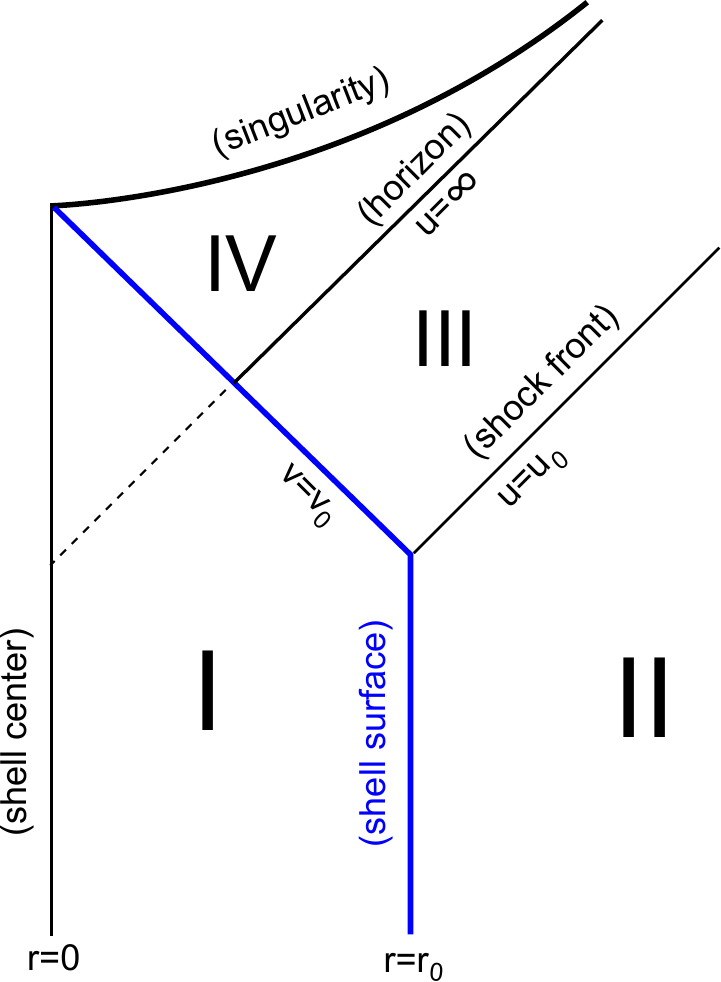}  \label{fig:collapse}
  }
\vspace{-15pt}
\end{wrapfigure}

In their early investigation of the fate of perturbing fields during gravitational collapse, De La Cruz, Chase, and Israel \cite{delacruz-chase-israel} introduced the null shell collapse spacetime pictured in figure \ref{fig:collapse}.  The spacetime is constructed by joining flat spacetime (region I) to Schwarzschild spacetime (regions II,III,IV) on a spherical hypersurface (blue line) that is initially tangent to the direction of time symmetry (i.e., stationary), but at a later time suddenly becomes tangent to the ingoing radial null direction (collapsing at the speed of light).  Following the standard junction conditions \cite{israel,poisson-book}, the induced metric is required to match on the shell, and the jump in extrinsic curvature is interpreted as the stress-energy of the shell.  This spacetime models a sudden and rapid gravitational collapse of a shell of matter.  In the words of the authors of \cite{delacruz-chase-israel}, the model is ``highly artificial from an astrophysicist's point of view, but does not violate any of the principles of relativity theory.''

We work in coordinates $(t,r,\theta,\phi)$ with $\partial_t$ the direction of time symmetry, $r$ the areal coordinate for symmetry two-spheres, and $(\theta, \phi)$ spherical coordinates on those spheres.  We take the collapse to begin at coordinate time $t=t_0$.  For $t<t_0$, the metric is that of the well-studied static spherical shell, given in our coordinates by
\begin{equation}
t<t_0: \qquad ds^2 = \begin{cases}
- (1-\tfrac{2M}{r_0})dt^2 + dr^2 + r^2 d\Omega^2 , & \quad r<r_0 \\
- (1-\tfrac{2M}{r}) dt^2 + (1-\tfrac{2M}{r})^{-1} dr^2 + r^2 d\Omega^2 , & \quad r>r_0 ,
\end{cases}
\end{equation}
where $d \Omega^2 = d \theta^2 + \sin^2 \theta d\phi^2$ is the metric of the two-sphere.  Here $r_0>2M$ is the initial coordinate radius of the shell.  However, we will take $r_0$ to be very large,
\begin{equation}
r_0 \gg M,
\end{equation}
so that for $t<t_0$ the spacetime is simply Minkowski spacetime everywhere.

On this Minkowski spacetime we place a point charge $q$ and a point magnetic dipole $\mu$.  The magnetic dipole is located at the origin and is oriented in the $z$ ($\theta=0$) direction, while the charge is located on the $x$ axis ($\phi=0$) at a distance $D<2M$ from the origin.  (We require $D<2M$ so that the charge ends up inside the black hole.)  In a suitable gauge the non-zero components of the vector potential are
\begin{align}
A_t & = q \sum_{\ell=0}^{\infty} \sum_{m=-\ell}^{\ell} \frac{r_<^\ell}{r_>^{\ell+1}} \frac{4\pi}{2\ell+1}\bar{Y}^{\ell m}(\tfrac{\pi}{2},0) Y^{\ell m}(\theta,\phi) \label{charge-field} \\
A_\phi & = \frac{\mu}{r} \sin^2\theta, \label{dipole-field}
\end{align}
where we have expanded the point charge field $A_t$ in spherical harmonics $Y^{\ell m}$ (with the conventions of \cite{jackson}).  Here $r_< (r_>)$ denotes the lesser (greater) of $r$ and $D$, and an overbar denotes complex conjugation.  By assumption, equations \eqref{charge-field} and \eqref{dipole-field} give the fields for $t<t_0$.  However, by causality they must also give the fields in the entirety of regions I and II, and, by continuity of the electromagnetic field, can therefore provide characteristic\footnote{Characteristic initial data refers to data on the union of two intersecting null surfaces, in this case the surfaces $u=u_0$ and $v=v_0$.} initial data for an evolution into the remaining portions of the spacetime.  Since our main concern is with the momentum radiated to infinity, we confine the evolution to region III, the Schwarzschild exterior.  We use double-null coordinates, related to Schwarzschild coordinates $(t,r)$ by
\begin{align}
u &= t-r_* \label{defineu} \\
v &= t+r_*, \label{definev}
\end{align}
where $r_*=r+2M \log (r/(2M)-1)$ is the tortoise coordinate.  The two-sphere at the onset of collapse (meeting point of regions I,II,III) is described by $u=u_0\equiv t_0-r_*(r_0)$ and $v=v_0\equiv t_0+r_*(r_0)$.   For $t>t_0$, $v=v_0$ is the collapsing shell's surface, while $u=u_0$ represents a spherical shock front propagating outward at the speed of light, bringing news of the sudden onset of collapse.  The (future) event horizon is located at $u\rightarrow \infty$ (fix $v$), while $v \rightarrow \infty$ (fix $u$) represents future null infinity ${\mathscr I}$, the vantage-point of a distant observer receiving the radiation.  The shell, shock front, horizon, and null infinity bound our evolution region III,  $u_0<u<\infty$ and $v_0 < v <\infty$.  The initial data is given on the shell and shock front, $v=v_0$ and $u=u_0$.

\subsection{Field Equations}

To solve the time-dependent Maxwell equation on the Schwarzschild background, we will use the master function approach \cite{cunningham-price-moncrief-1,cunningham-price-moncrief-2} as formulated in \cite{haas}.\footnote{We find two sign errors in \cite{haas}.  The first term in equation (2.11) should have a minus sign, while equation (2.12) should not have a minus sign.}  The vector potential is expanded into four scalar amplitudes, $A_a^{\ell m}$, $v^{\ell m}$ and $\tilde{v}^{\ell m}$,
\begin{align}
A_a & = \sum_{\ell=0}^{\infty} \sum_{m=-\ell}^{\ell} A_a^{\ell m} Y^{\ell m} \\
A_A & = \sum_{\ell=1}^{\infty} \sum_{m=-\ell}^{\ell} \left( v^{\ell m} Z_A^{\ell m} + \tilde{v}^{\ell m} X_A^{\ell m} \right),
\end{align}
where indices $a,b,c,...$ run over the coordinates $t,r$ and indices $A,B,C...$ run over the coordinates $\theta,\phi$.  Here $Z_A^{\ell m}=\partial_A Y^{\ell m}$ is the even parity vector spherical harmonic, while $X_A^{\ell m} = \epsilon_A{}^B \partial_B Y^{\ell m}$ is the odd parity vector spherical harmonic.  (The Levi-Civita tensor $\epsilon_{AB}$ is that of the unit two-sphere ($\epsilon_{\theta \phi}=\sin \theta$), and the index is raised with the metric of the unit two-sphere.)  One then defines master functions $\psi^{\ell m}_\pm$ in terms of the amplitudes by
\begin{align}
\psi_+^{\ell m} & = r^2 \left( \partial_t A_r^{\ell m} - \partial_r A_t^{\ell m} \right) \\
\psi_-^{\ell m} & = \ell(\ell+1)\tilde{v}^{\ell m},
\end{align}
and, using Maxwell's equations, it turns out that these functions entirely determine the field strength by
\begin{align}
F_{tr} & = \sum_{\ell=0}^{\infty} \sum_{m=-\ell}^{\ell} \frac{\psi^{\ell m}_+}{r^2} Y^{\ell m} \\
F_{tA} & = \sum_{\ell=1}^{\infty} \sum_{m=-\ell}^{\ell} \frac{1}{\ell(\ell+1)} \left\{\partial_t \psi_-^{\ell m} X_A^{\ell m} + (1-\tfrac{2M}{r})\partial_r \psi_+^{\ell m} Z_A^{\ell m}  \right\} \\
F_{rA} & = \sum_{\ell=1}^{\infty} \sum_{m=-\ell}^{\ell} \frac{1}{\ell(\ell+1)} \left\{ \partial_r \psi_-^{\ell m} X_A^{\ell m} + (1-\tfrac{2M}{r})^{-1} \partial_t \psi_+^{\ell m} Z_A^{\ell m} \right\} \\
F_{\theta \phi} & = \sum_{\ell=1}^{\infty} \sum_{m=-\ell}^{\ell} \psi_-^{\ell m} \sin \theta Y^{\ell m} .
\end{align}
Back in terms of double-null coordinates, Maxwell's equations imply that the master functions satisfy
\begin{equation}
\frac{\partial^2 \psi_{\pm}}{\partial u \partial v} + \left( 1-\frac{2M}{r} \right) \frac{\ell(\ell+1)}{4 r^2} \psi_{\pm} = 0 \qquad (\ell > 0).\label{waveeqn}
\end{equation}
Equation \eqref{waveeqn} applies only to the modes with $\ell>0$; the $\ell=0$ mode is non-radiative and requires a separate treatment.  (The non-radiative character follows immediately from Gauss' law.  Since our interest is in the radiation, we do not consider this mode.\footnote{Our initial solution for $\ell=0$, $A_t \propto 1/r$, is also a solution in Schwarzschild spacetime (when areal coordinates are identified, as they are in our shell metric).  This observation allows one to see explicitly that the $\ell=0$ mode does not evolve when the shell begins to collapse.})  Note that equation \eqref{waveeqn} is the same for all $m$ and each choice of $+$ or $-$.  The initial data for our evolution is provided by the static solutions, equations \eqref{charge-field}-\eqref{dipole-field}.  Since $D<2M$, the initial data surfaces $u=u_0$ and $v=v_0$ satisfy $r>D$, and we require the point charge solution only for $r>D$ (i.e., $r_>=r, r_<=D$).  Computing the master functions associated with the initial data then gives
\begin{align}
\psi^{\ell m}_+ & = 4 \pi q \frac{\ell+1}{2\ell+1} \left( \frac{D}{r} \right)^\ell \bar{Y}^{\ell m}(\tfrac{\pi}{2},0) \label{charge-initial-data} \\
\psi^{\ell m}_- & = \begin{cases} \sqrt{\frac{16\pi}{3}} \frac{\mu}{r}  , & \textrm{if } \ell=1 \textrm{ and } m=0 \\ 0, & \textrm{otherwise} \end{cases}.\label{dipole-initial-data}
\end{align}
Here we have used the fact that $X_{\phi}^{10}=\sqrt{3/(4\pi)} \sin^2\theta$.

Equations \eqref{charge-initial-data}-\eqref{dipole-initial-data} provide initial data on $u=u_0$ and $v=v_0$ for the evolution of equation \eqref{waveeqn}.  In order to avoid redundant integrations over different $m$-values (which satisfy the same wave equation), it is convenient to introduce real amplitudes $f^\ell_{\pm}$ such that
\begin{align}
\psi^{\ell m}_+ & = f^{\ell}_+ \frac{4\pi}{2\ell+1} \bar{Y}^{\ell m}(\tfrac{\pi}{2},0) \\
\psi^{10}_{-} & = f^1_- \sqrt{\tfrac{16\pi}{3}} .
\end{align}
The amplitudes satisfy the same wave equation,
\begin{equation}
\frac{\partial^2 f_{\pm}^{\ell}}{\partial u \partial v} + \left(1-\frac{2M}{r}\right) \frac{\ell(\ell+1)}{4 r^2} f_{\pm}^{\ell} = 0 \qquad (\ell>0), \label{waveeqnf}
\end{equation}
with initial values given by
\begin{align}
f^{\ell}_{+}|_{u=u_0,v=v_0} & = q (\ell+1)\left(\frac{D}{r}\right)^\ell \label{fplusinitial} \\
f^{1}_{-}|_{u=u_0,v=v_0} & = \frac{\mu}{r}. \label{fminusinitial}
\end{align}

\subsection{Radiated Energy and Momentum}

Our main interest in this evolution is the total energy and (especially) momentum radiated away.  At future null infinity, $v \rightarrow \infty$, the spacetime is flat and we may compute the total flux of energy and momentum using expressions valid in Minkowski spacetime.  Let $t^\alpha$ and $(\hat{x}_i)^\alpha$ ($i=1,2,3$) be the time and space translation Killing fields of flat spacetime.  (In Cartesian Minkowski coordinates, we have $t^\alpha=(1,0,0,0)$ and $(\hat{x}_i)^\mu=\delta^\mu{}_i$.)  These vector fields, when dotted into a conserved stress-energy tensor, give the associated current of energy and momentum, which can be integrated over the surface $v \rightarrow \infty$ to give the total radiated energy and ($i^{\textrm{th}}$ component of) momentum.  For the total electromagnetic energy and momentum lost through future null infinity in our spacetime, we have
\begin{align}
\Delta E & = \lim_{v \rightarrow \infty} \int \left( \tfrac{1}{2} T^\alpha{}_{\beta} t^\beta \partial_\alpha v \right) r^2 d\Omega d u \label{deltaEgen} \\
\Delta P_i & = - \lim_{v \rightarrow \infty} \int \left( \tfrac{1}{2} T^\alpha{}_{\beta} (\hat{x}_i)^\beta \partial_\alpha v \right) r^2 d\Omega du, \label{deltaPgen}
\end{align}
where $\theta,\phi$ range over the two-sphere (area element $d\Omega=\sin \theta d\theta d\phi$) and $u$ runs from $-\infty$ to $\infty$.  (However, for our field configuration the flux vanishes for $u<u_0$ and one may begin the integral at $u=u_0$.)  Here $T_{\mu \nu}$ is the stress-energy tensor of the electromagnetic field,
\begin{equation}
4 \pi T_{\mu \nu} = F_{\mu \alpha} F_{\nu}{}^{\alpha} - \tfrac{1}{4} g_{\mu \nu} F_{\alpha \beta} F^{\alpha \beta}.
\end{equation}

The radiated energy takes an elegant form in terms of the master functions; using the orthogonality of the spherical harmonics it is straightforward to show that
\begin{equation}
\Delta E = -\frac{1}{4\pi} \sum_{\ell=1}^{\infty} \sum_{m=-\ell}^{\ell} \frac{1}{\ell(\ell+1)} \lim_{v \rightarrow \infty} \int \left( |\partial_u \psi^{\ell m}_+|^2 + |\partial_u \psi^{\ell m}_-|^2 \right) d u, \label{DeltaE}
\end{equation}
where the vertical bars denote complex norm.  In deriving this equation and analogous flux equations below, we have used the fact that as $v \rightarrow \infty$ the general solution of equation \eqref{waveeqnf} is the sum of a pure function of $u$ and a pure function of $v$.  In terms of our real amplitudes, we have
\begin{equation}
\Delta E = - \lim_{v \rightarrow \infty} \int \left( \sum_{\ell=1}^{\infty} \frac{(\partial_u f^\ell_+)^2}{\ell(\ell+1)(2\ell+1)} + \tfrac{2}{3} (\partial_u f_-^1)^2 \right) d u.\label{DeltaEf}
\end{equation}

The general expression for the radiated momentum (analogous to equation \eqref{DeltaE} for the energy) will be less elegant and more laborious to derive, since it will involve Clebsch-Gordon-type coupling coefficients.  (Each spatial Killing vector $(\hat{x}_i)^{\alpha}$ is $\ell=1$, so that triple products of spherical harmonics appear under the integral.)  However, from the symmetry of the problem and the ``addition of angular momentum'' rules, it is straightforward to establish the following: (a) the radiated momentum in the $x$ direction takes the form of a coupling between even parity modes with $\ell$-values separated by 1; (b) the radiated momentum in the $y$ direction will come entirely from the $\ell=1$ even-parity mode multiplied by the $\ell=1$ odd-parity mode; and (c) the radiated momentum in the $z$ direction is vanishing.  Evaluating the angular integrals explicitly for low values of $\ell$, we have
\begin{align}
\Delta P_x & = - \lim_{v \rightarrow \infty} \int \left( \tfrac{1}{15} (\partial_u f_+^1) (\partial_u f_+^2) + \tfrac{2}{105} (\partial_u f_+^2) (\partial_u f_+^3) + \dots \right) du \label{DeltaPxf} \\
\Delta P_y & = - \lim_{v \rightarrow \infty} \int \left( \tfrac{1}{3} (\partial_u f^1_-) (\partial_u f^1_+) \right) du \label{DeltaPyf} \\
\Delta P_z & = 0. \label{DeltaPzf}
\end{align}
The radiated momentum in the $x$ direction would persist in the absence of the magnetic dipole and can be thought of as resulting from the asphericity of the collapse of a shell onto an off-center point charge.  This asphericity will result in a slight asphericity of the emitted radiation, leading to a kick.  On the other hand, the radiation in the $y$ direction requires both the dipole and the charge and may be thought of as due to a release of the initial field momentum (entirely in the $y$ direction) during the rapid collapse event.  Since the initial data for each real amplitude $f^\ell_{\pm}$ scales as $(D/r)^\ell$, we can anticipate that this will be the dominant contribution, i.e., $|P_y|>|P_x|$.  (We find that $|P_y|\approx 16 qD^2/(\mu M) |P_x|$.)  Furthermore, it is clear that we lose little accuracy by dropping terms higher order in $\ell$, represented by the $\dots$ in equation \eqref{DeltaPxf}.  (We find that the second term is smaller than the first by a factor of about $75$ for order unity parameters.)  It is clear that we may therefore neglect the third and higher terms, already not displayed in equation \eqref{DeltaPxf}.

\subsection{Parameter Scalings}

Some further simplification is helpful for numerical purposes and for revealing the scaling of the results with the parameters of the model.  Since $f_+^1$ and $f_-^1$ are proportional initially (see equations \eqref{fplusinitial} and \eqref{fminusinitial}) and satisfy the same linear wave equation \eqref{waveeqnf}, they are in fact proportional for all time.  We may thus keep track of both $\ell=1$ amplitudes by a single fiducial amplitude $f$.  We will define $f$ to be the numerical function obtained by integrating the wave equation \eqref{waveeqnf} with $\ell=1$ and $M=1$ and with initial data given by $f=1/r$ (on $u=u_0$ and $v=v_0$ in the limit of a large shell, $u_0 \rightarrow -\infty$).  Using equations \eqref{waveeqnf}-\eqref{fminusinitial}, the momentum flux formula \eqref{DeltaPyf} becomes
\begin{equation}\label{deltaPyI}
\Delta P_y = -\frac{\mu q D}{M^3} I ,
\end{equation}
where we define the fiducial integral $I$ by
\begin{equation}\label{fiducialI}
I = \lim_{v \rightarrow \infty} \int \tfrac{2}{3} (\partial_u f)^2 du,
\end{equation}
with the factor $2/3$ introduced for convenience.  The scaling of $\Delta P_y$ with $\mu$, $q$, and $D$ is evident from equations \eqref{fplusinitial} and \eqref{fminusinitial}, while the factor of $M^{-3}$ could have been guessed on dimensional grounds.\footnote{To establish equation \eqref{deltaPyI} it is helpful to introduce coordinates $\bar{u}=u/M$ and $\bar{v}=v/M$, which agree with $u$ and $v$ in the fiducial case $M=1$.}  The integral $I$ is simply a number that may be computed numerically.  From equation \eqref{fiducialI} we have that $I$ is manifestly positive, so that the kick is guaranteed to be in the direction of the initial hidden momentum, as expected.  We find numerically that $I \approx 1.35 \times 10^{-3}$ (see section \ref{sec:numerical}).  By conservation of momentum, the net kick delivered to the final black hole is
\begin{equation}\label{kickvel}
v^{\textrm{kick}}_y = \Delta P_y / M = - \frac{\mu q D}{M^4} I.
\end{equation}

From equations \eqref{pem}-\eqref{pmech}, the initial field configuration has momentum $P^{EM}_y = \mu q / D^2$ (counterbalanced by hidden mechanical momentum $P^{\textrm{mech}}_y=-\mu q / D^2$).  We may define the efficiency $\epsilon_{\textrm{kick}}$ of the hidden momentum kick as the fraction of this field momentum that is radiated away during the collapse.  From equation \eqref{deltaPyI} we have
\begin{equation}\label{efficiency}
\epsilon_{\textrm{kick}} = \frac{\Delta P_y}{P^{EM}_y} = \left( \frac{D}{M} \right)^3 I.
\end{equation}
Thus for separation distances $D$ of order the mass $M$ of the shell, the efficiency is about $0.1\%$.  For the maximum allowed separation distance, $D=2M$, the efficiency is $1.0\%$.

If we keep only the $\ell=1$ terms in the energy flux, then  equation \eqref{DeltaEf} shows that the fiducial integral $I$ also controls the energy flux,
\begin{equation}\label{DeltaEI}
\Delta E|_{\ell=1} = - \frac{\mu^2 + q^2 D^2}{M^3} I.
\end{equation}
We find numerically that the $\ell=2$ term is a correction of fractional order $10^{-3}(D/M)$, while the $\ell=3$ term is a correction of fractional order $10^{-5}(D/M)$.  Thus equation \eqref{DeltaEI} should give an excellent approximation to the energy flux.

Finally note that we may divide equations \eqref{deltaPyI} and \eqref{DeltaEI} to obtain
\begin{equation}
\frac{\Delta P_y}{\Delta E|_{\ell=1}} = \frac{\mu q D}{\mu^2+q^2 D^2} \label{pony}.
\end{equation}
The fiducial integral has canceled out, and we have a fully analytic result for the ratio of the $y$-momentum flux to the dominant energy flux.  We see that the ratio is controlled by a symmetric combination of the initial magnetic dipole moment $\mu$ and electric dipole moment $qD$, reaching a maximum of $1/2$ when the moments are equal.  Equation \eqref{pony} could also have been derived directly from equations \eqref{DeltaEf} and \eqref{DeltaPyf} using the fact that $f_+^1=(2qD/\mu)f_-^1$.

To summarize, we first presented our problem in terms of a general master-function formulation of Maxwell's equations on a Schwarzschild background.  Then we reduced that prescription to one involving real amplitudes $f^\ell_{\pm}$, whose initial values are given in equations \eqref{fplusinitial} and \eqref{fminusinitial}.  Once equation \eqref{waveeqnf} is integrated with these initial values, the fluxes of energy and momentum are given by equations \eqref{DeltaEf}-\eqref{DeltaPzf}.  Finally, we introduced a fiducial $f$, corresponding to integrating equation \eqref{waveeqnf} with $M=1$ and initial data $1/r$.  This integration should be carried out with initial shell radius taken to be very large, $u_0 \rightarrow -\infty$.  The integral $I$, defined by equation \eqref{fiducialI} in terms of this $f$, then contains all the information about the dominant energy flux $\Delta E|_{\ell=1}$ (equation \eqref{DeltaEI}) and momentum flux $\Delta P_y$ (equation \eqref{deltaPyI}) for any choice of model parameters $q,\mu,D,M$.

\subsection{Numerical Implementation}\label{sec:numerical}

\begin{figure}
    \includegraphics[width=.95\textwidth]{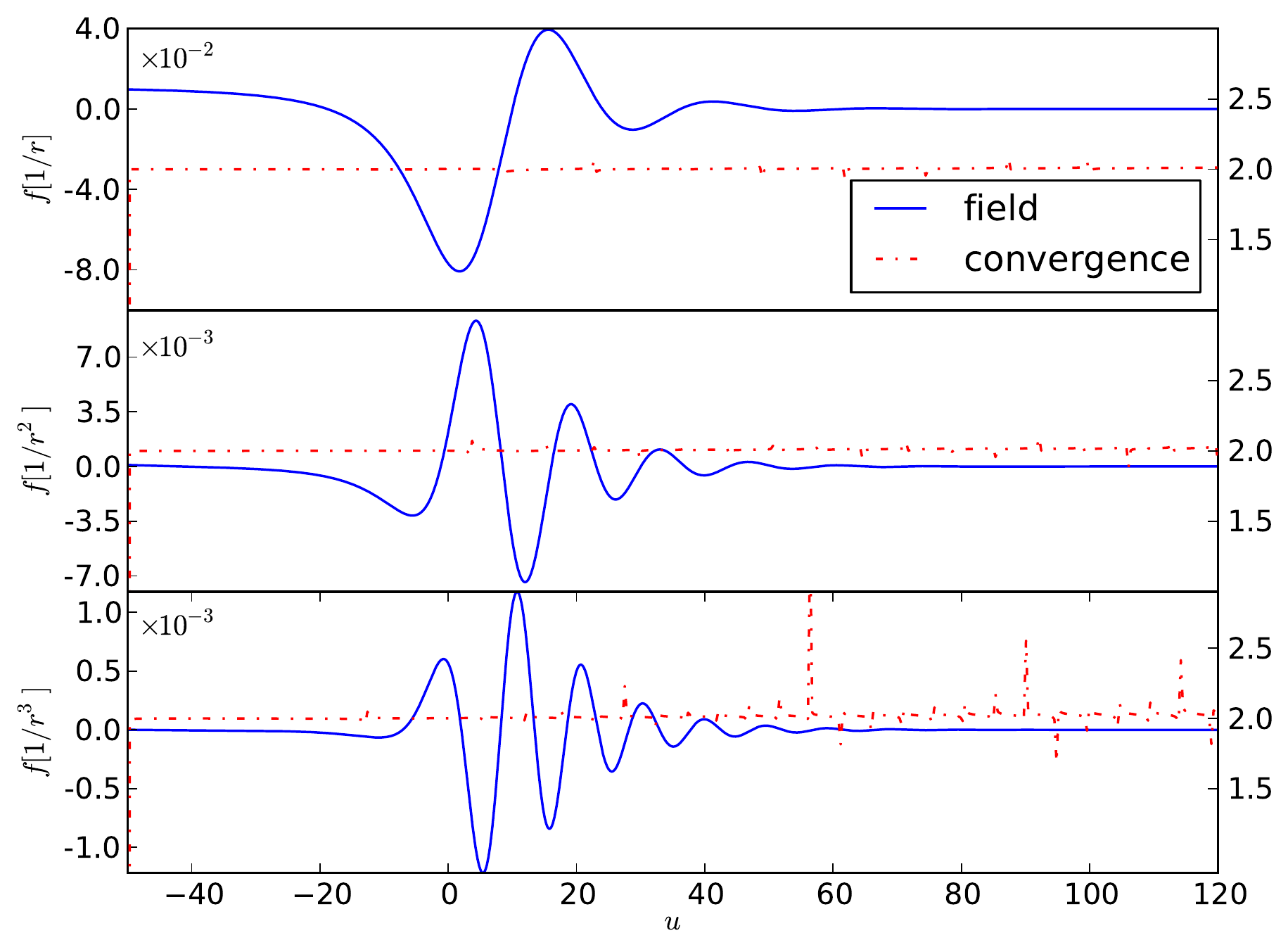}
    \caption{The field $f(u,v_{max})$ as a function of $u$ on the grid boundary $v=v_{max}$ which approximates ${\mathscr I}$.  The field (left scale) is evolved for the three different cases $\ell=1,2,3$.  The dashed line shows the convergence factor $c\approx 2$ (right scale) and demonstrates good 2nd order convergence.}
    \label{fig:field}
\end{figure}

We evolve equation \eqref{waveeqnf} using a 2nd order finite difference discretization,
\begin{equation}
f_N=f_E+f_W-f_S - h^2 V_c \frac{1}{2} \left(f_E+f_W\right) ,
\end{equation}
where the indices $N, S, W, E$ refer to north, south, west, east on the grid as in reference \cite{thornburg,gomez-winicour}, $h$ is the grid spacing, and $V_c$ is approximated by $V_c\approx (1/2) (V_E+V_W)$ and $V=(1-2M/r) \ell(\ell+1)/(4r^2)$. We evolve equation \eqref{waveeqnf} with $M=1$ on a grid of $u\in [u_0,u_{\textrm{max}}]$ and $v\in[0,v_{\textrm{max}}]$ with initial data $1/r^{\ell}$ for $\ell=1,2,3$. In principle our model requires $u_0 = -\infty$, $u_{\textrm{max}}=+\infty$ and $v_{\textrm{max}}=+\infty$, and we settle on values $u_0=-50$, $u_{\textrm{max}}=150$ and $v_{max}=200$ as sufficiently large (see below for error discussion). Figure \ref{fig:field} shows the result of these evolutions with $f$ evaluated on $v_{\textrm{max}}$, approximating future null infinity ${\mathscr I}$.  We performed each simulation for three different resolutions with grid spacing of $h=0.1, 0.2, 0.4$. The dashed line shows the convergence factor $c=\log_2\left((f_{4h}-f_{2h})/(f_{2h}-f_h)\right)$, which should be $c=2$ for 2nd order convergence.  Notice that the waves reach null infinity ($v=v_{\textrm{max}}$) around $u=-10$, peak around $u=+10$, and decay by about $u=60$. Since our shell is at $v=0$, these retarded times correspond (respectively) to emission from $r_*=\{5,-5,-30\}$, or $r \approx 4.53M$ (initial emission), $r \approx 2.05M$ (peak of emission) and $r \approx 2.00M$ (end of emission).  Thus the vast majority of the emission comes from the strong-field regime.

The $y$ momentum flux and the dominant energy flux are both captured by the fiducial integral $I$, equation \eqref{fiducialI}, which is computed from the $\ell=1$ evolution.  We find
\begin{equation}
I = 1.35 \times 10^{-3}.
\end{equation}
In establishing this result we have quantified three sources of error: finite resolution ($e_{\textrm{res}}=10^{-7}$), finite extraction radius  ($e_{v_{\textrm{max}}}=10^{-5}$), and finite initial shell radius ($e_{u_0}=3\times 10^{-5}$).  Our resolution error estimate is done via Richardson extrapolation.  For the finite extraction radius error, we vary $v_{\textrm{max}}$ and fit a function of the form $I_0+I_1/v_{\textrm{max}}$ in $1/v_{\textrm{max}}$.  (The grid is kept square, so that $u_{\textrm{max}}$ varies as well.)  For the initial shell radius error we similarly fit a function of the form $I_0 + I_1/u_0$ in $1/u_0$. The value $I$ quoted above is extrapolated in $v_{\textrm{max}}$, but not in resolution $h$ or $u_0$. From the $\ell=2$ and $\ell=3$ evolutions we also compute the higher order corrections to the radiated energy, equation \eqref{DeltaEf}, and the radiated $x$ momentum, equation \eqref{DeltaPxf}.  For order unity parameters, the subleading energy term is a fractional correction of order $10^{-2}$, while the $x$ momentum flux is smaller than the $y$ momentum flux by a factor of order $10$.  The text below equation \eqref{DeltaPzf} gives the precise numbers, along with parameter scalings.

As a further check of our code we reproduce the results of the numerical run done in reference \cite{delacruz-chase-israel}.  In particular, for their parameter choices of $M=1/2$ and $u_0=-10$, we are able to reproduce their integral $I_1$ (see text below their equation (2)) to the accuracy given in the paper.

\section{Discussion}\label{sec:discussion}

In this paper we have explored the effect of dynamical processes on hidden momentum in one particular scenario involving gravitational collapse.  The main result may be presented as either a kick velocity of the final black hole or an efficiency of the conversion of hidden momentum to ordinary momentum by the collapse (see equations \eqref{kickvel}-\eqref{efficiency}),
\begin{equation}\label{hare}
|v_y^{\textrm{kick}}| \approx (400\, \textrm{km}/\textrm{s})\, \frac{\mu q D}{M^4}, \qquad \epsilon_{\textrm{kick}} \approx 0.14\% \left( \frac{D}{M} \right)^3.
\end{equation}
For ease of eventual comparison to purely gravitational phenomena, we measure the kick in kilometers per second, while leaving $\mu,q,D,M$ in natural units.  The efficiency of the process scales with the ratio of the size of the charge-and-dipole configuration, $D$, to the final black hole size $R=2M$.  If the mass $M$ of the shell is very large, so that the eventual black hole is much larger than the charge-and-dipole, then the kick becomes small.  This stands to reason, since the black hole engulfs both the hidden momentum within the dipole and the (equal and opposite) field momentum stored outside, leading to zero net kick.  On the other hand, when the size of the charge-and-dipole approaches the size of the final black hole, the black hole swallows only a portion of the field momentum, leading to a net kick.  We see no a priori way to estimate the percentage swallowed, which in principle could have been negligible.  Our numerical calculations have determined it to be about $0.1\%$, which leads to the prefactor of 400 km/s when the model parameters are expressed in Gaussian geometrized units.  The $\mu q D/M^4$ factor will be very small for ordinary laboratory and astrophysical systems (see section \ref{sec:gen} for some discussion), and we are not attempting to apply our model to presently observable electromagnetic phenomena.  Rather, we are primarily concerned with exploring hidden momentum in a dynamical context, and our particular calculations show that its sudden destruction can lead to a kick.

\subsection{Application to Gravity}

A second goal of the paper is to use the electromagnetic problem to help elucidate the mechanism behind the bobbing-and-kicks behavior seen in binary black hole mergers.  To compare directly to the kicks observed in binary black hole simulations we must make several extrapolations.  First, we must use the analogy between gravitation and electromagnetism, involving $q \sim m$ and $\mu \sim S$, where $m$ and $S$ are the mass and spin of a black hole in the (comparable mass, comparable spin) binary.  Second, since the shell of mass $M$ enforces our merger and produces a final black hole of mass $\sim M$, we should also take $M \sim m$ to approximate the merger of black holes of mass $\sim m$.  Third, to approximate the superkicks premerger configuration, we should take the distance between the particles to also be of order the mass/size of the black holes, $D \sim m$, and set the spin to be of order the maximal value, $S\sim m^2$.  Finally, we must extrapolate the kick formula \eqref{hare} beyond the test-Maxwell-field regime where it is strictly valid, since the above parameter identifications entail $q \sim M$ and $\mu \sim M^2$.  These extrapolations are crude, and a combination of order-unity errors could easily migrate the answer by an order of magnitude.  We therefore take the numerical value of 400 km/s only as a rough ballpark for the size of kicks that a hidden momentum kick mechanism could generate.  No such estimate was made in the original work proposing this mechanism \cite{gralla-harte-wald}, and our investigation could in principle have revealed the effect to be negligible.  That our figure is within an order of magnitude of the observed superkicks of 4000 km/s gives plausibility---but by no means definitiveness---to the hidden momentum kick mechanism.

We therefore believe that the hidden momentum kick mechanism may have a role to play in explaining the superkick phenomenon, and could 
in principle account for the entire effect.  However, an alternative proposal \cite{pretorius} also predicts kicks of the correct qualitative and roughly the correct quantitative character.  In this scenario one imagines that the bobbing motion of the black holes provides a small modulation to the dominant emission (due to the quasi-circular motion), which induces an asymmetry in that emission, resulting in the kick.  Since the bobbing motion is of direct relevance in producing the kick in this scenario, we will call this the ``motion scenario''.  This scenario does not rely on the presence of a merger to generate the kick, and a number of simulations of grazing collisions of black holes (i.e., near the threshold between capture/merging and scattering) \cite{sperhake-berti-cardoso-pretorius-yunes} have indicated that large kicks can be obtained even in the non-merging case.  However, in these simulations the black holes complete at least a full orbit before separating \cite{uli-comment}, and one could therefore still view the kick as a consequence of release of quasi-static field momentum due to the sudden separation of the bodies.  Thus the simulations significantly inform the discussion but do not conclusively distinguish between the two scenarios.

The work reported here helps further distinguish the scenarios by isolating the relevant physics of our scenario (a rapid event destroying the quasi-static configuration) and creating a situation where the relevant physics of the motion scenario is entirely absent.  Indeed, our ``binary'' consists of particles at \textit{rest}, with the emission entirely determined by the merger's destruction of the initially static field configuration.  That we can recover kicks in the ballpark of those seen in numerical simulations \textit{without any orbital motion at all} suggests that the rapid destruction of the quasi-static field configuration may indeed be playing an important role in producing the superkicks.  However, the uncertainty inherent in the extrapolation of our results to the gravitational case leaves plenty of room for the effect to be too small to fully account for the superkick.   The motion effect could be dominant, or both effects could be playing an important role.

The original paper on null-shell collapse in the presence of perturbing fields \cite{delacruz-chase-israel} treated a gravitational quadrupole in addition to an electromagnetic dipole.  This raises the question of why a purely gravitational calculation was not attempted here.  The answer is that the equivalence principle forbids a static binary in general relativity---all gravitating bodies attract.  A gravitational binary can only be maintained with orbital velocity, and so it is difficult to cleanly separate effects due to motion from those due to merger or collapse.  Studying an electromagnetic analog enables us to completely eliminate motion while preserving the other essential features of the pre-merger superkicks configuration.

\subsection{Hidden Momentum More Generally}\label{sec:gen}

We conclude with a brief discussion of other physical and astrophysical systems in which hidden momentum may have a role to play.  First, since the phenomenon is basically independent of the internal structure of the body (being enforced by the presence of quasi-static field momentum and the conservation of total momentum), similar kick phenomena are to be expected for mergers and near-merger scattering of neutron stars, boson stars, or any other object that could be simulated.  However, the kick will be smaller for bodies that are less compact.  Returning to the electromagnetic case where the notion of hidden momentum is most precisely defined, one could imagine that a magnetized neutron star moves through an external magnetic field, generating an ``external'' electric field in the local rest frame.  This would be an astrophysical example of the original hidden momentum setup, and our heuristic would predict a kick if the neutron star underwent collapse.  However, even assuming the strongest known magnetar moving through the field of another such object at an orbital separation of ten times the radius, the total hidden momentum corresponds to a kick of only about a millimeter per second.  Furthermore, plasma dynamics would likely be important for this system.

Having a large electromagnetic hidden momentum kick requires compact objects with strong intrinsic electric and magnetic fields.  One possible avenue for further exploration would be the addition of electric charge to the spinning black holes already considered, which generates an effective magnetic dipole moment as well.   In this case there should be significant hidden momentum of both the electromagnetic and gravitational variety.

\ack{We thank Ulrich Sperhake for helpful correspondence.  S.G. acknowledges support from NASA through the
Einstein Fellowship Program, Grant PF1-120082. Parts of this work were supported by NSF grant PHY1005632.}

\hspace{20pt}

\end{document}